# Magnetic structure of the kagome metal $YbFe_6Ge_6$ in view of Bragg diffraction

S. W. Lovesey

*ISIS Facility, STFC, Didcot, Oxfordshire OX11 0QX, United Kingdom*
*Diamond Light Source, Harwell Science and Innovation Campus, Didcot, Oxfordshire OX11 0DE, United Kingdom*
*Department of Physics, Oxford University, Oxford OX1 3PU, UK*

**Abstract** A material in possession of localized 4f-electron magnetism and delocalized 3d-electron or band magnetism can often present enigmatic physical phenomena, and there has been a longstanding interest in the kagome metal $YbFe_6Ge_6$. More recently, because of an investigation of a so-called anomalous Hall effect, or topological Hall effect, and magnetic neutron Bragg diffraction [W. Yao *et al.*, Phys. Rev. Lett. **134**, 186501 (2025)]. Iron moments in the two-dimensional layers of a hexagonal nuclear structure undergo collinear antiferromagnetic order below a temperature $\approx 500$ K. The moments depart from the c axis in a spontaneous transition at $\approx 63$ K to an orthorhombic structure. The magnetism of Yb ions appears to behave independently, which can be confirmed using resonant x-ray diffraction enhanced by a Fe atomic resonance. The inferred magnetic space group is a P(parity)T(time)-symmetric (anti-inversion $\bar{1}'$) collinear antiferromagnet. A linear magnetoelectric effect is allowed, as in historically important chromium sesquioxide, and Kerr rotation and the piezomagnetic effect are forbidden. Symmetry informed Bragg diffraction patterns for future x-ray and neutron experiments are shown to be rich in Fe magnetic properties of orthorhombic $YbFe_6Ge_6$, including space-spin correlations, anapoles and Dirac quadrupoles familiar in high-$T_c$ ceramic superconductors.

## I. INTRODUCTION

Ternary intermetallics $RFe_6Ge_6$ crystallize in closely related structures that result from insertion of the R element within the $Ge_8$ holes of the hexagonal host (P6/mmm) [1, 2, 3]. The magnetic properties are characterized by an antiferromagnetic ordering of the Fe sublattice above room temperature, and a stacking sequence $+ - + -$ along the hexagonal $c_h$-axis depicted in Fig. 1. Iron moments align with this axis, and moment values range between 1.5 $\mu_B$ and 2.2 $\mu_B$. There is zero molecular field on the R sites, and a concomitant low ordering temperature of the R sublattice caused by weak R–R interactions. In the compound of interest, $YbFe_6Ge_6$, resonant x-ray Bragg diffraction can distinguish Fe from Yb ($Yb^{3+}$, $4f^{13}$) magnetism by tuning the x-ray energy to an Fe atomic resonance. Iron moments align with the kagome plane in a spontaneous transition at $\approx 63$ K. They adopt a collinear antiferromagnetic structure depicted in Fig. 2, which belongs to an orthorhombic magnetic crystal class that includes anti-inversion ($\bar{1}'$) [4].

Anti-inversion among the elements of symmetry in a magnetic crystal class exerts a profound influence on diffraction amplitudes. For Bragg diffraction of x-rays enhanced by an atomic resonance, charge and magnetic contributions to scattering amplitudes are in phase. Thus, there is no interference between charge and magnetic contributions to Bragg diffraction patterns gathered from a compound that presents a linear magnetoelectric effect. Moreover, coupling to helicity in primary x-rays is forbidden, with no difference in the intensity of a Bragg spot observed with opposite handed x-rays. In contrast, anti-inversion invariance imposes a 90º phase shift between nuclear and magnetic contributions in neutron scattering, and they are in quadrature in the intensity of a Bragg spot. A corollary is that the classical polarized neutron diffraction technique, using a departure from unity of the ratio of intensities for primary neutron beams of opposite polarization, is not available for magnetoelectric compounds. In their measurement of the magnetization distribution in $Cr_2O_3$, Brown *et al.* exploited spherical neutron polarimetry [5].

Our symmetry informed diffraction patterns for the resonant x-ray [6] and magnetic neutron [7] diffraction by $YbFe_6Ge_6$ expose electronic correlations that can be confronted with simulations. Dipoles in resonant x-ray diffraction at the atomic K edge using a parity-even event are proportional to the orbital angular momentum of the resonant ion. Quadrupoles are charge-like (non-magnetic Templeton-Templeton scattering) and their angular anisotropy reveals the spatial symmetry of the crystal environment. Parity-odd and time-odd (Dirac) multipoles are compulsory in our diffraction amplitudes, because Fe ions use acentric positions in the magnetic structure [6, 7]. Diffraction patterns for the resonant x-ray and magnetic neutron diffraction by Fe ions in $YbFe_6Ge_6$ are predicted to include anapoles (Dirac dipoles), space-spin correlations, and Dirac quadrupoles familiar in high-$T_c$ ceramic superconductors [7]. To make the main body of the paper accessible many mathematical details reside in two appendices.

## II. CRYSTAL AND MAGNETIC STRUCTURES

Iron moments in $YbFe_6Ge_6$ depart from the c axis in P6/m′m′m′ (No. 191.241, Fig. 1 [8]) in a spontaneous transition at a temperature ≈ 63 K. Iron ions occupy acentric Wyckoff positions (8m) and (4l) in the orthorhombic space group Cm′mm, with cell dimensions $a \approx$ 8.828 Å, $b \approx$ 5.097 Å, $c \approx$ 8.092 Å [4], and a basis {(1, 2, 0), (−1, 0, 0), (0, 0, 1)} relative to hexagonal ($a_h$, $b_h$, $c_h$). Axial dipoles are always collinear in the (4l) position and allowed to be non-collinear in (8m). Reflection vectors for hexagonal and orthorhombic structures are labelled ($H_o$, $K_o$, $L_o$) and ($h$, $k$, $l$), respectively, with $h = H_o + 2K_o$, $k = −H_o$, $l = L_o$. The orthorhombic structure belongs to the centrosymmetric magnetic crystal class m′mm, and anti-inversion ($\bar{1}'$) is among the elements of symmetry (identical to properties of 6/m′m′m′ in Fig 1). Kerr rotation [9] and the piezomagnetic effect are not permitted by m′mm. In the following analysis of diffraction patterns, Fe multipoles are referred to orthogonal vectors labelled (ξ, η, ζ) derived from the orthorhombic unit cell depicted in Fig. 2. Table I contains the discrete symmetries of multipoles in amplitudes for resonant x-ray and magnetic neutron diffraction [6, 7].

# III. RESONANT X-RAY DIFFRACTION

Our resonant x-ray diffraction amplitudes are derived from universal expressions [6] and an electronic structure factor Eq. (A1). In line with current practice, primary photon polarizations labelled σ and π are perpendicular and parallel to the plane of scattering, respectively, and secondary polarization carry a prime as in Fig. 3. Regarding diffraction amplitudes, (π'σ) and (σ'σ) apply to primary σ polarization rotated to the π channel and returned to the σ channel, respectively. In the diffraction setting crystals are rotated around the reflection vector by an angle ψ (an azimuthal angle scan). In contrast to neutron diffraction, the fixed x-ray energy limits the number of Bragg spots available in a measured pattern.

Valence states accessed by photo-ejected electrons interact with neighbouring ions when x-rays excite a core resonance. Thus, electronic multipoles in the ground state observed in diffraction are rotationally anisotropic with a symmetry corresponding to the position symmetry of the resonant ion. Absence conditions in Bragg diffraction can be violated by relatively weak spots arising from non-spherical atomic charge (Templeton-Templeton scattering) [10, 11]. Tuning the energy of x-rays to an atomic resonance has two obvious benefits. In the first place, there is a welcome enhancement of Bragg spot intensities and, secondly, spots are element specific. There are four scattering amplitudes labelled by photon polarization, two with unrotated and two with rotated states of polarization [6, 10]. Strong Thomson scattering, by spherically symmetric atomic charge, that overwhelms weak signals is absent in rotated channels of polarization. It is allowed in unrotated channels of polarization using a parity-even absorption, but absent in a parity-odd absorption, e.g., electric dipole - electric dipole (E1-E1) and electric dipole - electric quadrupole (E1-E2) events. The range of values of the multipole rank K is fixed by the triangle rule, and K = 0 - 2, K = 1 - 3 and K = 0 - 4 for E1-E1, E1-E2 and E2-E2 events, respectively.

Cell dimensions of YbFe$_6$Ge$_6$ are too small to execute resonant x-ray Bragg diffraction enhanced by iron L$_{2,3}$ edges at a photon energy E ≈ 0.71 keV (photon wavelength ≈ (12.4/E) Å with E in units of keV). The iron K edge with E ≈ 7.115 keV gives access to 4p (E1) and 3d (E2) valence states [11].

Parity-even ($\sigma_\pi$ = +1) diffraction amplitudes E1-E1 and E2-E2 incorporate a time signature $\sigma_\theta (-1)^K$ = +1 [6], whereupon electronic structure factors Eqs. (A2) and (A3) are zero, to an excellent approximation, for Miller indices odd $l$ and even K. The corresponding (σ′σ) amplitude for an E1-E1 event is zero, because it does not contain dipoles K = 1, unlike (σ′σ) for an E2-E2 event [6]. Amplitudes for Wyckoff positions (8m) and (4l) are identical for even $k$ apart from a numerical factor. For a reflection (0, $k$, $l$) with $k$ = 2$n$, $l$ = (2$m$ + 1) and an E1-E1 absorption event,

$$(\pi'\sigma) = 8 (-1)^{m+n} \langle T^1_{+1} \rangle'' [\sin(\theta) \cos(\beta) - \cos(\theta) \sin(\beta) \sin(\psi)], \qquad (1)$$

with $\cos(\beta) = -k/[k^2 + (bl/c)^2]^{1/2}$. The Fe dipole moment parallel to the $a_h$-axis, $\langle T^1_{+1} \rangle'' \propto \langle T^1_\eta \rangle$, is observed. For absorption at a K edge, axial dipoles are proportional to the orbital angular momentum of the resonant ion. The azimuthal angle $\psi$ measures rotation of the crystal sample about the reflection vector $(0, k, l)$, and the orthorhombic $\xi$ axis and the y axis in Fig, 3 are parallel for $\psi = 0$.

Templeton-Templeton scattering in an E1-E1 event is provided by the quadrupoles $\langle T^2_Q \rangle$ with projections $Q = 0, \pm 2$. Setting $l = 0$ and using odd $h$, $k$ with $(h + k) = 2n$ in $(h, k, 0)$, for example, one finds,

$$(\pi'\sigma)(8m) = 8 (-1)^n \sin(\psi) \langle T^2_{+2} \rangle'' [\sin(\theta) \cos(\psi) \sin(2\chi) - \cos(\theta) \cos(2\chi)].$$

$$(\pi'\sigma)(4l) = \sin(\theta) \sin(2\psi) [\sqrt{6} \langle T^2_0 \rangle + 2 \cos(2\chi) \langle T^2_{+2} \rangle']$$

$$+ 4 \cos(\theta) \sin(\psi) \sin(2\chi) \langle T^2_{+2} \rangle'. \qquad (2)$$

Diffraction amplitudes (8m) and (4l) are proportional to $\sin(\psi)$, but otherwise very different functions of of the azimuthal angle. The relations $\Psi^K_Q(8m) = -\Psi^K_{-Q}(8m)$ and $\Psi^K_Q(4l) = \Psi^K_{-Q}(4l)$ for odd $k$ are responsible for the differences. Thus, $(\pi'\sigma)(8m)$ is conventional Templeton-Templeton scattering expected at a forbidden reflection with odd $h$. In Eq. (2), $\cos(\chi) = -[h/\sin(\theta)](\lambda/2a)$, and $\zeta$ and z axes are parallel at $\psi = 0$. Amplitudes $(\pi'\sigma)(8m)$ and $(\pi'\sigma)(4l)$ for $(h, k, 0)$ and even $h$, $k$ possess identical structures, namely, $(\pi'\sigma)(4l)$ in Eq. (2).

Wyckoff positions (8m) and (4l) are acentric and Dirac multipoles $\langle G^K_Q \rangle$ in Table I are compulsory, and they contribute in diffraction using a parity-odd E1-E2 absorption event. Operator equivalents for $\langle G^K_Q \rangle$ are presented in Ref. [12]. With $\sigma_\pi = \sigma_\theta = -1$, structure factors Eqs. (A2) and (A3) are zero for odd $l$. Notably, $\Psi^K_{-Q}(8m) = (-1)^{K+k} \Psi^K_Q(8m)$, whereas $\Psi^K_{-Q}(4l) = (-1)^K \Psi^K_Q(4l)$. Anapoles, or Dirac dipoles, $\langle \mathbf{G}^1 \rangle$ contribute to bulk properties ($h = k = l = 0$) with $\Psi^1_{+1}(8m) \propto \langle G^1_{+1} \rangle'$, and likewise for $\Psi^1_{+1}(4l)$. Notably, anapoles contribute to diffraction in the unrotated channel $(\sigma'\sigma)$; no magnetic multipoles contribute to the same channel for diffraction enhanced by an E1-E1 event. Using odd $h$, $k$ with $(h + k) = 2n$ in $(h, k, 0)$,

$$(\sigma'\sigma)(8m) = (8/5) \sqrt{6} (-1)^n \cos(\chi) \cos(\theta) \cos(\psi) [\langle G^1_{+1} \rangle'' + (1/3)\sqrt{5} \langle G^2_{+1} \rangle'] + ... \qquad (3)$$

Contributions to $(\sigma'\sigma)$ are truncated at the level of quadrupoles, and contributions from two remaining octupoles $\langle G^3_{+1} \rangle''$ and $\langle G^3_{+3} \rangle''$ are available from Ref. [6]. As expected, amplitudes $(\sigma'\sigma)(8m)$ and $(\sigma'\sigma)(4l)$ are not the same; e.g., the former contains $\langle G^1_{+1} \rangle'' \propto \langle G^1_\eta \rangle$ and the latter contains $\langle G^1_{+1} \rangle' \propto \langle G^1_\xi \rangle$. Specifically,

$$(\sigma'\sigma)(4l) = -(4/5)\sqrt{6}\sin(\chi)\cos(\theta)\cos(\psi)[\langle G^1_{+1}\rangle' - (1/3)\sqrt{5}\langle G^2_{+1}\rangle''] + \ldots, \quad (4)$$

and, like $(\sigma'\sigma)(8m)$, it is an even function of $\psi$. The factor $[\cos(\theta)\cos(\psi)]$ in Eqs. (3) and (4) is replaced by $[\sin(2\theta)\sin(\psi)]$ in the amplitude in the rotated channel of polarization $(\pi'\sigma)$, and it contains one additional quadrupole derived from $\langle G^2_{+1}\rangle = [\langle G^2_{+1}\rangle' + i\langle G^2_{+1}\rangle'']$.

### IV. MAGNETIC NEUTRON DIFFRACTION

A dependence of the magnetic neutron scattering amplitude on both the magnitude and direction of the reflection vector $\kappa$ is a most valuable property of the technique. It enables the measurement of the magnetization density, or its spatial Fourier transform more correctly. The dependence of amplitudes on the magnitude of the reflection vector appears in radial integrals illustrated in Fig. 4. In particular, radial integrals $(h_1)$ and $\langle j_2(\kappa)\rangle$ are derived from spherical Bessel functions of orders 1 and 2, respectively, and the maximum of $\langle j_2(\kappa)\rangle$ occurs at $\kappa \approx 6$ Å$^{-1}$. Dirac and axial quadrupoles are $\langle g^2 \rangle \propto [(h_1)\{\mathbf{S} \otimes \mathbf{n}\}^2]$ and $\langle t^2 \rangle \propto [\langle j_2(\kappa)\rangle \{(\mathbf{S} \times \mathbf{n})\,\mathbf{n}\}]$ [7]. The cited operator equivalents are reviewed in Appendix B. Suffice to mention that they engage electronic variables spin ($\mathbf{S}$), position ($\mathbf{n}$), and the Zel'dovich anapole ($\mathbf{S} \times \mathbf{n}$). Higher-order axial $\langle t^K \rangle$ ($\sigma_\theta = -1$, $\sigma_\pi = +1$) and polar $\langle g^K \rangle$ ($\sigma_\theta = \sigma_\pi = -1$) multipoles are not included in the present study. Intensity of a Bragg spot is $|\langle \mathbf{Q}_\perp \rangle|^2 = |\langle \mathbf{Q} \rangle|^2 - |(\mathbf{e} \cdot \langle \mathbf{Q} \rangle)|^2$, with a unit vector $\mathbf{e} = (\boldsymbol{\kappa}/\kappa)$. In general, $\langle \mathbf{Q} \rangle = [\langle \mathbf{Q} \rangle^{(+)} + \langle \mathbf{Q} \rangle^{(-)}]$, where superscripts label axial $^{(+)}$ and polar $^{(-)}$ intermediate amplitudes. The dipole approximation for Bragg diffraction derived by Schwinger $\langle \mathbf{Q} \rangle^{(+)} = (1/2)\langle \boldsymbol{\mu} \rangle$ uses $\langle j_0(\kappa)\rangle = 1$ for $\kappa \to 0$, and a magnetic moment $\langle \boldsymbol{\mu} \rangle = \langle 2\mathbf{S} + \mathbf{L} \rangle$.

Neutron scattering amplitudes for Wyckoff positions (8m) and (4l) are derived from electronic structure factors Eqs. (A2) and (A3), respectively. Amplitudes are purely imaginary, as expected for a PT-symmetric magnetic crystal class. Projections Q are odd integers since all multipoles are time-odd. In consequence, the Miller index $l = (2m + 1)$ for axial $(\sigma_\pi = +1)$ magnetic multipoles, cf. measured diffraction patterns reported in Fig. S18 [4], and $l = 2m$ for polar $(\sigma_\pi = -1)$ multipoles. The condition $(h + k) = 2n$ satisfies centring in the magnetic space group.

Wyckoff position (8m) and even $k$, with a common factor $i8\,(-1)^{m+n}$.

$$\langle Q_\xi \rangle^{(+)} \approx -e_\xi\, e_\eta \langle t^2_{+1}\rangle', \quad \langle Q_\eta \rangle^{(+)} \approx -(3/\sqrt{2})\langle t^1_{+1}\rangle'' + (e_\xi^2 - e_\zeta^2)\langle t^2_{+1}\rangle', \quad (5)$$

$$\langle Q_\zeta \rangle^{(+)} \approx e_\eta\, e_\zeta \langle t^2_{+1}\rangle'.$$

For odd $k$,

$$\langle Q_\xi \rangle^{(+)} \approx -(3/\sqrt{2})\langle t^1_{+1}\rangle' - (e_\eta^2 - e_\zeta^2)\langle t^2_{+1}\rangle'', \quad \langle Q_\eta \rangle^{(+)} \approx -e_\xi\, e_\eta \langle t^2_{+1}\rangle'', \quad (6)$$

$$\langle Q\zeta \rangle^{(+)} \approx e_\xi\, e_\zeta\, \langle t^2_{+1}\rangle''.$$

Observed axial dipoles $\langle \mathbf{t}^1 \rangle$ are parallel to the $\eta$- and $\xi$-axes, respectively. Quadrupoles $\langle t^2_{+1}\rangle = [\langle t^2_{+1}\rangle' + i\langle t^2_{+1}\rangle'']$ significant at large reflection vectors $\kappa \approx 6$ Å$^{-1}$ are seen to contribute to all diffraction amplitudes.

Wyckoff position (4l) and all $k$, and a common factor $i4\,(-1)^{m+k}$. Otherwise, $\langle \mathbf{Q}\rangle^{(+)}$ is identical to (8m) with even $k$.

Turning to polar multipoles one needs even $l = 2m$. The anapole $\langle \mathbf{g}^1 \rangle$ is represented by Cartesian components of $\langle \mathbf{D} \rangle$ defined in Eq. (B3). Observed anapoles are $\langle D\xi \rangle$ and $\langle D\eta \rangle$ for even and odd $k$, respectively. For Wyckoff positions (8m) and even $k$, with a common factor $i8\,(-1)^{m+n}$,

$$\langle Q_{\perp,\xi}\rangle^{(-)} \approx 2\, e_\xi\, e_\eta\, e_\zeta\, \langle g^2_{+1}\rangle'', \quad \langle Q_{\perp,\eta}\rangle^{(-)} \approx e_\zeta\, [\langle D\xi\rangle + (2e_\eta^2 - 1)\, \langle g^2_{+1}\rangle''],$$

(7)

$$\langle Q_{\perp,\zeta}\rangle^{(-)} \approx e_\eta\, [-\langle D\xi\rangle + (2e_\zeta^2 - 1)\, \langle g^2_{+1}\rangle''].$$

In particular, $\langle \mathbf{Q}_\perp \rangle^{(-)} = (0, 0, \langle Q_{\perp,\zeta}\rangle^{(-)})$ for $l = 0$. The corresponding results for odd $k$ are,

$$\langle Q_{\perp,\xi}\rangle^{(-)} \approx e_\zeta\, [-\langle D\eta\rangle + (2e_\xi^2 - 1)\, \langle g^2_{+1}\rangle'],\quad \langle Q_{\perp,\eta}\rangle^{(-)} \approx 2\, e_\xi\, e_\eta\, e_\zeta\, \langle g^2_{+1}\rangle',$$

(8)

$$\langle Q_{\perp,\zeta}\rangle^{(-)} \approx e_\xi\, [\langle D\eta\rangle + (2e_\zeta^2 - 1)\, \langle g^2_{+1}\rangle'].$$

Notably, Dirac quadrupoles $\langle g^2_{+1}\rangle''$ and $\langle g^2_{+1}\rangle'$ are present in all components of the polar amplitudes. Amplitudes for Wyckoff position (4l) and all $k$, have a common factor $i4\,(-1)^{m+k}$. Otherwise, $\langle \mathbf{Q}_\perp \rangle^{(-)}$ is identical to (8m) amplitudes in Eq. (7) with even $k$.

## V. CONCLUSIONS

There is currently no direct evidence that the magnetic structure of the kagome metal YbFe$_6$Ge$_6$ is PT-symmetric [4]. Neutron polarization analysis was not used by Yao *et al.* to verify the symmetry. Furthermore, the distribution of Fe magnetization and the Fe atomic form factor in YbFe$_6$Ge$_6$ is work to be done. Shull has published these measurements for elemental iron, cf. Fig 2 in Ref. [13].

I report symmetry informed calculations of neutron and resonant x-ray diffraction patterns for the collinear AFM structure Cm'mm implied by limited neutron Bragg diffraction patterns [4]. A linear magnetoelectric effect is allowed by Cm′mm, as with historically important chromium sesquioxide, and Kerr rotation [9] and the piezomagnetic effect are not permitted.

PT-symmetry in the magnetic structure protects against coupling to neutron polarization and, also, circular polarization in the primary x-ray beam. In consequence, the method for neutron polarization analysis used by Shull is not available, but the distribution of iron magnetization can be measured with precision using neutron polarimetry. Shull's data on elemental Fe extends to $s = \sin(\theta)/\lambda = 1.157$ Å$^{-1}$, and data on PT-symmetric $Cr_2O_3$ extends to $s = 0.75$ Å$^{-1}$ [5]. Quadrupoles that are predicted here to exist in the distribution of Fe magnetization are caused by correlations between electronic anapole and position variables. They have optimal values at a smaller wavevector $s \approx 0.50$ Å$^{-1}$ ($\kappa \approx 6$ Å$^{-1}$) [Eqs. (5) & (6)] that can be confronted with simulations of the electronic structure [14]. Dirac quadrupoles in neutron diffraction amplitudes [Eqs. (7) & (8)] offer a sound interpretation of magnetic neutron diffraction by high-$T_c$ materials [7]. The Dirac quadrupole operator equivalent is made from spin and position variables [7, 16]. Results for the radial integral ($h_1$) presented in Fig. 4 indicate that Fe Dirac quadrupoles possess an optimal value for $s \approx 0.16$ Å$^{-1}$. Axial dipoles, closely related to the magnetic moment Eq. (B2), and anapoles Eq. (B3) contribute to neutron scattering amplitudes presented in Eqs. (5)-(8).

Resonant x-ray Bragg diffraction enhanced by a Fe absorption event has the obvious advantage of distinguishing properties of Fe and Yb ions in $YbFe_6Ge_6$. No such experiments are in the literature. Exact scattering amplitudes reported in §III exploit universal expressions for E-E1 and E1-E2 amplitudes [6]. Parity-even enhancement E2-E2 at the iron K edge measures orbital angular momentum that is different for $3d^5$ and $3d^6$ atomic configurations; the former is spherically symmetric and the additional electron in the latter configuration is exposed to the electron-lattice interaction. Relevant E2-E2 diffraction amplitudes are readily calculated from available universal expressions [6]. A null unrotated (σ′σ) parity-even (E1-E1) amplitude is predicted. Eqs. (1) & (2) for the rotated channel of polarization demonstrate a dependence of amplitudes on a magnetic dipole and charge-like quadrupoles (Templeton-Templeton scattering). Anapoles are a main attraction in Eqs. (3) & (4) for Dirac amplitudes (σ′σ) and an E1-E2 event, with Bragg spots predicted to be free of stronger E1-E1 signals. Amplitudes in Eqs. (1)-(4) include the dependence on an azimuthal angle that is dictated by Fe symmetry of the Wyckoff position [15].

**ACKNOWLEDGEMENT** Dr W. Yao endorsed IRs ($\Psi_{15} + \Psi_{16}$) in Table S4 for the low-temperature magnetic structure of $YbFe_6Ge_6$, and not IRs ($\Psi_{17} - \Psi_{18}$) stated in the text [4]. Professor R. D. Johnson spotted incorrect IRs, and Dr D. D. Khalyavin confirmed the magnetic space groups and prepared Figs. 1 and 2. Professor G. van der Laan generated Figs. 3 and 4.

# APPENDIX A: ELECTRONIC STRUCTURE FACTOR

A universal spherical structure factor of rank K,

$$\Psi^K_Q = [\exp(i\boldsymbol{\kappa}\cdot\mathbf{d})\,\langle O^K_Q\rangle_\mathbf{d}], \tag{A1}$$

determines x-ray and neutron Bragg diffraction amplitudes. The generic electronic multipole $\langle O^K_Q\rangle$ possesses $(2K+1)$ projections in the interval $-K \le Q \le K$, and the complex conjugate obeys $(-1)^Q \langle O^K_{-Q}\rangle = \langle O^K_Q\rangle^*$. Our phase convention for real and imaginary parts labelled by single and double primes is $\langle O^K_Q\rangle = [\langle O^K_Q\rangle' + i\langle O^K_Q\rangle'']$. Cartesian dipole moments in an orthorhombic cell $(\xi, \eta, \zeta)$ are $\langle O^1_\xi\rangle = -\sqrt{2}\,\langle O^1_{+1}\rangle'$, $\langle O^1_\eta\rangle = -\sqrt{2}\,\langle O^1_{+1}\rangle''$, and $\langle O^1_\zeta\rangle = \langle O^1_0\rangle$. Multipole types encountered in resonant x-ray and magnetic neutron diffraction are reviewed in Table I.

Multipoles for resonant x-ray and neutron scattering abide by the same discrete symmetry requirements but they are different in detail. Magnetic neutron multipoles are time-odd, and axial or polar (Dirac) for an acentric Wyckoff position. Multipoles in resonant x-ray scattering are parity-even for electric dipole-electric dipole (E1-E1) and electric quadrupole-electric quadrupole (E2-E2) absorption events. They are time-even (time-odd) for even (odd) rank K. Parity-odd multipoles are permitted for acentric Wyckoff positions and they are either time-even or time-odd. The structure factor $\Psi^K_Q$ is informed of all elements of symmetry in the magnetic space group. In Eq. (A1), the reflection vector $\boldsymbol{\kappa}$ is defined by integer Miller indices $(h, k, l)$ with $(h + k) = 2n$ from a centring condition. The implied sum is over Fe ions in sites $\mathbf{d}$ in a magnetic unit cell. In more detail, Eq. (A1) possesses information about the relevant Wyckoff positions available in the Bilbao table MWYCKPOS for the magnetic symmetry of interest [8]. Site symmetry that might constrain projections Q is given in the same table. Wyckoff positions in a unit cell are related by operations listed in the table MGENPOS [8]. Taken together, the two tables provide all information required to evaluate Eq. (A1) and, thereafter, x-ray and neutron Bragg diffraction amplitudes [6, 7].

Evaluated for Wyckoff positions (8m) in the magnetic space group Cm′mm (No. 65.483),

$$\Psi^K_Q(8m) = 2\,(-1)^n\,[\exp(i\varphi) + \sigma_\pi(-1)^Q \exp(-i\varphi)]$$

$$\times [\langle O^K_Q\rangle + \sigma_\pi(-1)^k(-1)^{K+Q}\langle O^K_{-Q}\rangle], \tag{A2}$$

where $\varphi = 2\pi l z$. We use a general coordinate $z = 1/4$ [4]. In Eq. (A2), $\sigma_\pi$ and $\sigma_\theta$ are signatures of discrete symmetries of parity and time, respectively, with $\sigma_\pi = +1\,(-1)$ for axial (polar) and $\sigma_\theta = +1\,(-1)$ for time-even (time-odd, magnetic). Site symmetry (8m) demands $\sigma_\theta(-1)^Q = +1$ and no more. Site symmetry for Wyckoff positions (4l) is sturdier with a structure factor,

$$\Psi^K_Q(4l) = 2\,(-1)^k \langle O^K_Q \rangle\,[\exp(i\varphi) + \sigma_\pi (-1)^Q \exp(-i\varphi)]. \qquad (A3)$$

The electronic multipole in Eq. (A3) satisfies $\langle O^K_Q \rangle = \sigma_\theta \sigma_\pi (-1)^K \langle O^K_{-Q} \rangle$ in addition to $\sigma_\theta (-1)^Q = +1$. Multipoles for parity-even E1-E1 and E2-E2 absorption events possess a time signature $\sigma_\theta (-1)^K = +1$.

## APPENDIX B: NEUTRON OPERATOR EQUIVALENTS

An intermediate operator **Q** for magnetic neutron scattering by electrons can be taken to be [7],

$$\mathbf{Q} = \exp(i\mathbf{R}_j \cdot \boldsymbol{\kappa})\,[\mathbf{s}_j - (i/\hbar\kappa^2)\,(\mathbf{e} \times \mathbf{p}_j)]. \qquad (B1)$$

Here, $\mathbf{R}_j$, $\mathbf{s}_j$, and $\mathbf{p}_j$ are operators for electron position, spin and linear momentum, respectively. The unit vector $\mathbf{e} = (\boldsymbol{\kappa}/\kappa)$. The amplitude of magnetic neutron Bragg diffraction $\langle \mathbf{Q}_\perp \rangle = [\mathbf{e} \times (\mathbf{Q} \times \mathbf{e})]$ yields an intensity $|\langle \mathbf{Q}_\perp \rangle|^2 = |\langle \mathbf{Q} \rangle|^2 - |(\mathbf{e} \cdot \langle \mathbf{Q} \rangle)|^2$. Note that **Q** is arbitrary to within a scalar function proportional to the reflection vector $\boldsymbol{\kappa}$.

The dipole approximation for Bragg diffraction derived by Schwinger is $\langle \mathbf{Q} \rangle^{(+)} \approx (1/2) \langle \boldsymbol{\mu} \rangle$ with a magnetic moment $\langle \boldsymbol{\mu} \rangle = \langle 2\mathbf{S} + \mathbf{L} \rangle$ [7]. A useful approximation to the axial dipole $\langle \mathbf{t}^1 \rangle$ uses a gyromagnetic ratio g for an orbital contribution $\langle \mathbf{L} \rangle \propto (g-2)\langle \mathbf{S} \rangle$,

$$\langle \mathbf{t}^1 \rangle \approx (\langle \boldsymbol{\mu} \rangle/3)\,[\langle j_0(\kappa) \rangle + \langle j_2(\kappa) \rangle\,(g-2)/g]. \qquad (B2)$$

The radial integral $\langle j_0(\kappa) \rangle$ is illustrated in Fig. 4. Even rank multipoles arise from correlations between anapole and position $\mathbf{n} = \mathbf{R}/R$ variables that are probed in the orbital-spin contribution to $\langle \mathbf{Q} \rangle^{(+)}$. Specifically, a quadrupole operator equivalent $\mathbf{t}^2 \propto [\langle j_2(\kappa) \rangle\,\{(\mathbf{S} \times \mathbf{n})\,\mathbf{n}\}]$, where $(\mathbf{S} \times \mathbf{n})$ is the operator for a spin anapole. As for the radial integral, $\langle j_2(0) \rangle = 0$ and the maximum of $\langle j_2(\kappa) \rangle$ occurs at $\kappa \approx 6$ Å$^{-1}$.

In general $\langle \mathbf{Q} \rangle = [\langle \mathbf{Q} \rangle^{(+)} + \langle \mathbf{Q} \rangle^{(-)}]$, where a polar contribution $\langle \mathbf{Q} \rangle^{(-)}$ is essential for magnetic ions at acentric Wyckoff positions. A polar dipole $\langle \mathbf{D} \rangle$ depends on three radial integrals illustrated in Fig. 4. We use,

$$\langle \mathbf{D} \rangle = (1/2)\,[\,i(g_1)\,\langle \mathbf{n} \rangle + 3\,(h_1)\,\langle \mathbf{S} \times \mathbf{n} \rangle - (j_0)\,\langle \boldsymbol{\Omega} \rangle], \qquad (B3)$$

and an orbital anapole $\langle \boldsymbol{\Omega} \rangle = [\langle \mathbf{L} \times \mathbf{n} \rangle - \langle \mathbf{n} \times \mathbf{L} \rangle]$. Radial integrals $(g_1)$ and $(j_0)$ diverge in the forward direction of scattering ($\kappa \to 0$). We appeal to the orbital-spin contribution to $\langle \mathbf{Q} \rangle^{(-)}$ for an approximation to the polar quadrupole. It is proportional to $(h_1)$ at the chosen level of working, as in Eq. (B3), and arrive at,

$$\langle \mathbf{Q} \rangle^{(-)} \approx i(\mathbf{e} \times \langle \mathbf{D} \rangle) - i\sqrt{3}\,\{\mathbf{e} \otimes \langle \mathbf{H}^2 \rangle\}^1, \qquad (B4)$$

with a quadrupole operator equivalent,

$$\mathbf{H}^2 = \sqrt{5}\,(h_1)\{\mathbf{S} \otimes \mathbf{n}\}^2. \tag{B5}$$

Tensor products in Eqs. (B4) and (B5) follow a standard definition for two spherical tensors of ranks a and b, namely,

$$\{A^a \otimes B^b\}^K_Q = A^a{}_\alpha B^b{}_\beta\,(a\alpha b\beta|KQ), \tag{B6}$$

with a sum over repeated indices $\alpha, \beta$. The Clebsch-Gordan coefficient in Eq. (B6) can be defined in terms of a 3j-symbol,

$$(a\alpha b\beta|KQ) = (-1)^{-a+b-Q}\sqrt{2K+1}\begin{pmatrix} a & b & K \\ \alpha & \beta & -Q \end{pmatrix}. \tag{B7}$$

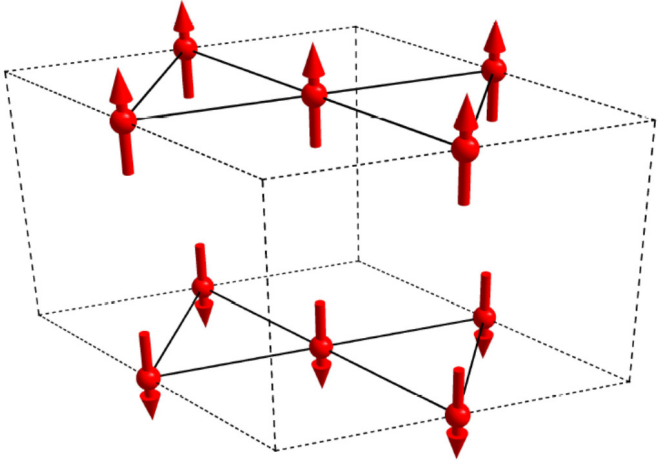

**FIG. 1**. Collinear antiferromagnetic order of iron dipole moments in YbFe$_6$Ge$_6$ at room temperature. Hexagonal magnetic space group P6/m′m′m′ (No. 191.241), and PT-symmetric magnetic crystal class 6/m′m′m′.

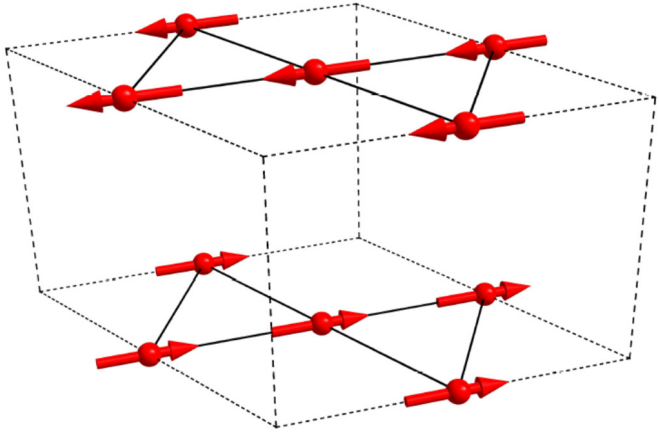

**FIG. 2**. PT-symmetric orthorhombic configuration of iron dipole moments in YbFe$_6$Ge$_6$ at a low temperature < 63 K. Magnetic space group Cm'mm (No. 65.483).

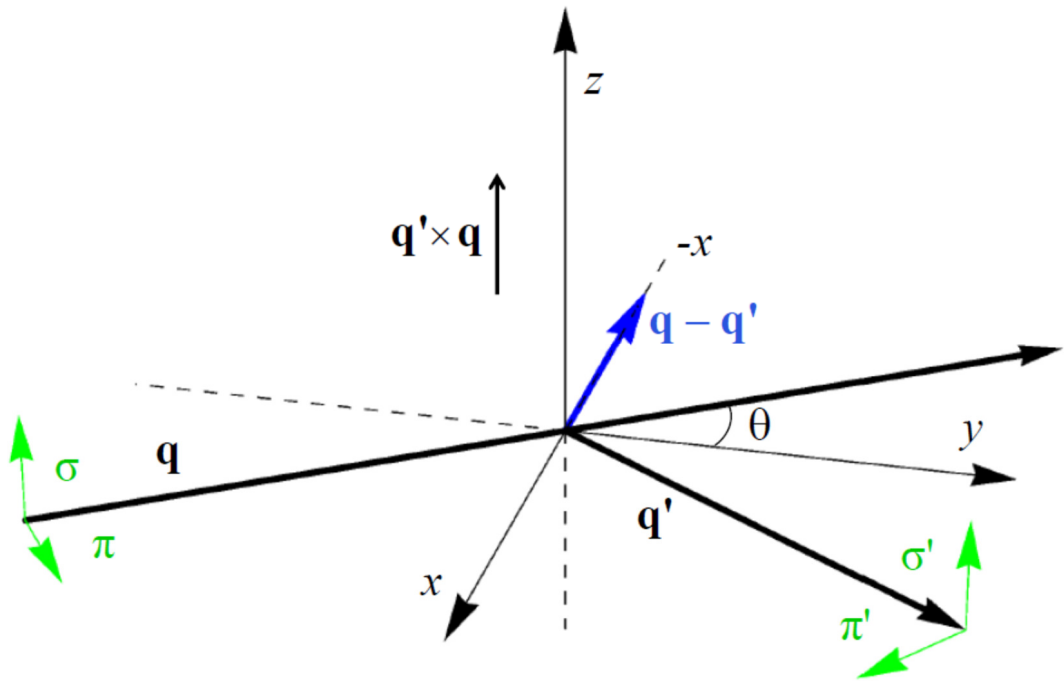

**FIG. 3**. Primary (σ, π) and secondary (σ′, π′) states of polarization. Corresponding wavevectors **q** and **q′** subtend an angle 2θ. The Bragg condition for diffraction is met when **q** − **q′** coincides with a reflection vector (h, k, l) of the orthorhombic reciprocal lattice. Crystal vectors that define local axes (ξ, η, ζ) and the depicted Cartesian (x, y, z) coincide in the nominal setting of the crystal.

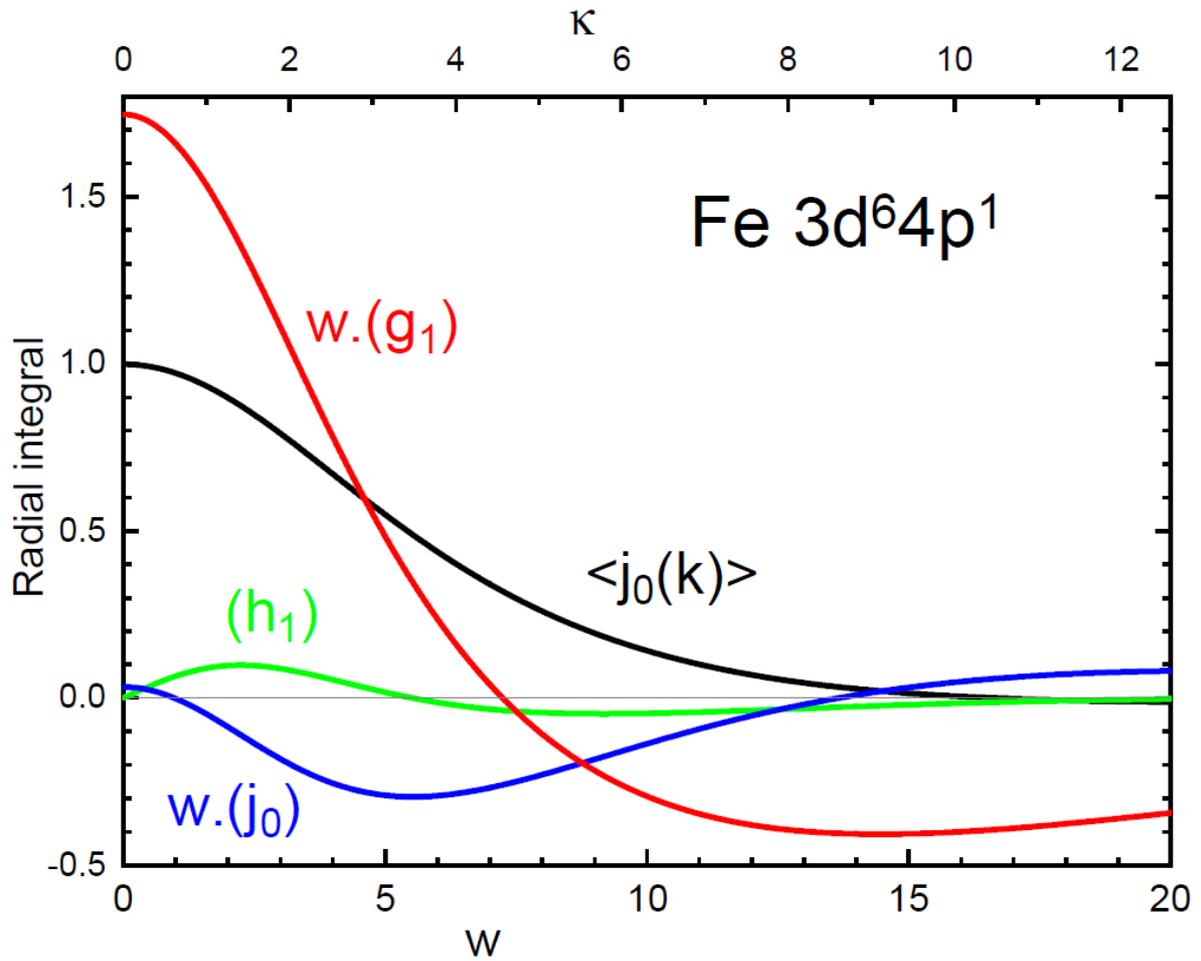

**FIG. 4**. Radial integrals for an Fe ion displayed as a function of the magnitude of the reflection vector $\kappa = 4\pi s$ with $s = \sin(\theta)/\lambda$ (Å$^{-1}$), Bragg angle $\theta$ as in Fig. 3, and neutron wavelength $\lambda$. Also, w = $3a_o\kappa$ where $a_o$ is the Bohr radius. Standard radial integral $\langle j_0(\kappa)\rangle$ in black appears in the axial dipole Eq. (B2). Red, green and blue curves radial integrals in the polar dipole Eq. (B3), with the orbital-spin radial integral (h$_1$) depicted in green. Radial integrals (g$_1$) and (j$_0$) diverge in the forward direction of scattering. Displayed quantities include a factor w, to give w.(g$_1$) and w.(j$_0$) that behave nicely for $\kappa \to 0$. Calculations and figure by G. van der Laan.

**TABLE I.** A generic spherical multipole $\langle O^K_Q \rangle$ has integer rank K and (2K + 1) projections Q in the interval $-K \leq Q \leq K$. Angular brackets $\langle ... \rangle$ denote the expectation value, or time average, of the enclosed spherical tensor operator. Parity ($\sigma_\pi$) and time ($\sigma_\theta$) signatures = ±1, e.g., $\langle t^K_Q \rangle$ for magnetic neutron diffraction is parity-even ($\sigma_\pi$ = +1) and time-odd ($\sigma_\theta$ = −1). Iron ions in YbFe$_6$Ge$_6$ occupy acentric Wyckoff positions, and Dirac multipoles $\langle g^K_Q \rangle$ and $\langle G^K_Q \rangle$ with ($\sigma_\theta \sigma_\pi$) = +1 are permitted. Operator equivalents are found in Refs. [7, 12].

| Signature | $\sigma_\pi$ | $\sigma_\theta$ |
|---|---|---|
| *Neutrons* | | |
| $\langle t^K_Q \rangle$ | +1 | −1 |
| $\langle g^K_Q \rangle$ | −1 | −1 |
| *Photons* | | |
| $\langle T^K_Q \rangle$ | +1 | $(-1)^K$ |
| $\langle G^K_Q \rangle$ | −1 | −1 |

**References.**